\documentclass[twocolumn,showpacs,preprintnumbers,amsmath,amssymb]{revtex4}

\usepackage{bbm}
\usepackage{epsfig}
\usepackage{amsfonts}
\usepackage{amsmath}
\usepackage{graphicx}
\usepackage{dcolumn}
\usepackage{bm}
\newtheorem{thm}{Theorem}[section]

\begin{document}


\title{Information Extraction from Decohering Systems via Indirect Continuous Measurement}

\author{Narayan Ganesan}
\email{ng@ese.wustl.edu}
\author{Tzyh-Jong Tarn}%
 \email{tarn@wuauto.wustl.edu}
\affiliation{%
Electrical and Systems Engineering\\
Washington University in St. Louis\\
}%


\begin{abstract}
In this article we discuss a general information extraction scheme to gain knowledge of the state and the amount
of decoherence based on indirect continuous measurement. The purpose of this information extraction is to
determine the feedback in order to control decoherence based on an ``output equation", described by a bilinear
form. An interacting bilinear form of control system is used instead of master equation to analyze the dynamic
properties of the system. The analysis of continuous measurement on a decohering quantum system has not been
extensively studied before.
\end{abstract}

\maketitle


Many authors have studied the application of feedback methods in control of
decoherence\cite{doherty},\cite{horoshko}. Technological advances enabling manipulation, control of quantum
systems and recent advances in quantum measurements using weak coupling, non-demolition principles etc, have
opened up avenues for employing feedback based control strategies for quantum systems
\cite{wiseman},\cite{horoshko},\cite{wallentowitz}.
%

In \cite{doherty} Doherty et al, studied hamiltonian feedback strategies for quantum systems under a POVM(Positive
Operator Valued Measurement) scheme. The authors proposed choosing a form for hamiltonian based on the output
obtained from generalized POVM measurements. Since it is not always practically feasible to realize any form for
the hamiltonian of the system at will or choose the type of measurement observable in POVM such strategies not
readily implementable. Viola et. al\cite{viola2} studied the group theoretic properties of the operators generated
by interacting hamiltonians and proposed an open loop strategy to nullify the effects of decoherence through
clever choice of controls. Recent developments by the same authors relaxed a few conditions regarding arbitrarily
strong and fast "bang-bang" pulses and the repertoire of available operators.

Continuous measurement by quantum probe\cite{itano} which stems from the theory of quantum Zeno effect can be used
to monitor a quantum system that might be coupled to extraneous unknown reservoirs. In this work we consider a
continuous time measurement scheme with the help of a probe, which is a quantum system interacting with the system
of interest. It is also shown that only the type of interaction hamiltonian $H_{SE}$ is sufficient to analyze the
decoherence properties of the system without delving too much into statistical properties of the environment/bath.
It is to be noted that instead of regulating the coherence between basis states - when the exact amount of
coherence changes dynamically during the course of quantum control or computation - we merely choose to render it
independent of certain undesirable interactions. The former has the shortcoming that the exact coherence
information to be regulated is not always known a priori as processing an unknown coherence (viz. processing
unknown quantum information) is the purpose of quantum computation. Whilst most of the previous work on
decoherence control mentioned above has been based on studying the Markovian master equation of the reduced
density matrix of the system we propose to use a systems theoretic approach to analyze the effects of decoherence.
Since the state of a decohering quantum system rapidly reduces to a mixed state from a pure state, the density
matrix approach and other stochastic dynamic approaches seems to be the best tool at hand. Unfortunately such
tools are not the most convenient when analyzing the controllability and other basic geometric properties of open
quantum systems, since the governing equations for the evolution is hard to rewrite in a bilinear form that is
most amenable to such analysis. Moreover the systems theoretic approach offers the elegance and important results
can be borrowed from the rich literature on non-linear control theory that could be applied to the model of
quantum control systems as discussed in \cite{htc}. As was previously shown by the
authors\cite{ganesan}\cite{ganesan2} the Lie Algebraic properties of the interaction hamiltonians can be exploited
to learn about and control the decoherence in an open quantum system by employing the so called {\it output
equation} that monitors the coherence between states. But the theory is sufficiently general to encompass other
forms of output such as non-demolition or just expected value of any real observable.

Consider an open quantum system interacting with the environment described by,
\begin{eqnarray*}
&\frac{\partial\xi(t,x)}{\partial t}=&[H_0\otimes
\mathcal{I}_e(t,x)+\mathcal{I}_s \otimes H_e(t,x)+H_{SE}(t,x)
\\&& +\sum_{i=1}^{r}u_i(t)H_i\otimes \mathcal{I}_e(t,x)]\xi(t,x) \label{opqusys}
\end{eqnarray*}
Here the argument $x$ denotes the spatial dependance of the combined system-environment state $\xi(t,x)$ as well
as control hamiltonians $H_i$, and where $u_i$ are the strength of the control respectively. $H_0, H_E, H_{SE}$
are the system, environment and interaction hamiltonian acting on $\mathcal{H}_s$, $\mathcal{H}_e$ and
$\mathcal{H}_s\otimes\mathcal{H}_e$ (system, environment and the combined) Hilbert spaces respectively. For ease
of notation we will suppress the spatial dependance. Define an output equation which could either be a
non-demolition measurement or a general bilinear form given by,
\begin{equation}
y(t)=\langle \xi(t)|C(t)|\xi(t)\rangle \label{output}
\end{equation}
where again $C(t,x)$ is assumed to be time-varying operator acting on the system Hilbert space. For instance for a
finite system the non-hermitian operator $C=|m\rangle\langle n|$ when plugged in eq. (\ref{output}) would yield
the coherence between the respective states of the system or for an electro-optic system the operator
$C=a\exp(i\omega t)+a^\dagger\exp(-i\omega t)$ would yield the output of a real non-demolition observation
performed on the system. In order to study the invariance properties with respect to the system dynamics of the
above time dependent quantum system, we define $f(t,x,u_1,\cdots,u_r,H_{SB})=y(t,\xi) \mbox{ for } t\in[t_0,t_f]$
to be a complex scalar map as a function of the control functions and the interaction Hamiltonian $H_{SB}$ over a
prescribed time interval. The function $f$ is said to be invariant or the signal $y(t,\xi)$ is said to decoupled
from the interaction Hamiltonian $H_{SB}$ if,
\begin{equation}
f(t,x,u_1,\cdots,u_r,H_{SB}) = f(t,x,u_1,\cdots,u_r,0) \label{cond9}
\end{equation}
for all admissible control functions $u_1,\cdots,u_r$ and a given interaction Hamiltonian $H_{SB}$. Then the
condition for such an output signal to be decoupled from the interaction hamiltonian in the open loop case is
given by the following theorem derived in\cite{ganesan}, which follows an iterative construction in terms of
system operators.
\begin{thm}
Let $\mathcal{C}_0=C(t)$ and for $n=1,2,\cdots$
\begin{align*}
&\tilde{C}_n=\mbox{span}\{ad^j_{H_i}\mathcal{C}_{n-1}(t)|j=0,1,\ldots;i=1,\ldots,r\}\\
&\mathcal{C}_n=\left\{ \left(ad_H+\frac{\partial}{\partial
t}\right)^j\tilde{C}_n; j=0,1,\cdots \right\}
\end{align*}
Define a distribution of quantum operators, $
\tilde{\mathcal{C}}(t)=\mbox{span}\{\mathcal{C}_1(t),\mathcal{C}_2(t),\cdots{},\mathcal{C}_n(t),\cdots{}\}$. The
output equation (\ref{output}) of the quantum system is decoupled from the environmental interactions if and only
if,
\[[\tilde{\mathcal{C}}(t), H_{SE}(t)]=0\]
\end{thm}
The same condition is relaxed to $[\tilde{\mathcal{C}}(t), H_{SE}(t)]\subset \tilde{\mathcal{C}}(t)$ with the
application of feedback control of the form $u(t)=\alpha(\xi(t))+\beta(\xi(t))v$, where $\alpha$ and $\beta$ are
analytic functions of the state, suggesting the usefulness of such feedback techniques in controlling decoherence.
It is to be noted that the feedback discussed here is classical and deterministic function of the state $\xi$,
which differs from the formulation of quantum feedback discussed in\cite{wiseman}. As a detailed scheme to compute
and implement feedback is itself discussed in\cite{ganesan2}, in this article we concern ourselves only with an
information extraction scheme which is key to implementing such feedback strategies. We do so via indirect
continuous measurement using quantum probe on a system {\it undergoing decoherence} to gain knowledge of the state
$\xi$ which according to authors' best knowledge has not been sufficiently studied before. In order to provide the
feedback another quantum system labeled as the {\it quantum probe} or simply {\it probe} which interacts
continuously with the system will be used to gather information. The Hamiltonian of the system now modifies as
$H_0+H_{SP}(t)+H_e+H_{SE}+ H_{PE} +\sum_{i=1}^{r}u_i(t)H_i$, where $H_{0} = H_S + H_P$, is the free hamiltonian of
the system and probe. The interaction hamiltonian between different subsystems are denoted by their respective
subscripts.
Decoherence, as discussed by Zurek\cite{zurek2} develops strong correlation between preferred pointer basis of a
system and the states of environment. Under the framework of continuous measurement the question arises as to
what can be said about the relationship between the pointer states of the system and the coupled (measurement)
device. As is expected, if the corresponding pointer states of system and probe are not naturally correlated
under the system probe interaction hamiltonian $H_{SP}$, the later would not be a good measuring
device\cite{zurek}. The pointer basis play an all too important role in the analysis of such systems.
%
We now analyze the expected measurement results of the probe observable which is weakly coupled to the decohering
system so as to gain knowledge about the state of the system. The expected measurement result is shown to reveal
information about the degree of decoherence of the system which could be used to design the feedback control. Let
$|s_i\rangle$ and $|A_i\rangle$ be the pointer basis of the system and probe respectively. The above mentioned
bases inherit all the qualities of a pointer basis\cite{zurek}, viz. orthogonal and be able to distinguish,
develop correlation between the bases under interaction etc. The probe observable that is measured by the
environment is $\hat{A} = \sum_{i=0}^{N-1} a_i |A_i\rangle \langle A_i |$ where $a_i$'s are eigenvalues of the
probe observable. The probe system interaction hamiltonian will follow the structure $H_{SP} =
g(t).\hat{s}.\hat{P}$ where $\hat{P}=\sum_{l=0}^{N-1} l|B_l\rangle \langle B_l|$ acts on probe Hilbert space and $
\hat{s}=\sum_{l=0}^{N-1} s_j|s_j\rangle \langle s_j|$ acts on system's Hilbert space respectively. The signal
$g(t)$ is the coupling strength which is assumed to be modulated externally. The operator $\hat{P}$ also generates
shift in the {$\hat{A_k}$} basis. Hence the basis {$|A_k \rangle$} and {$|B_l \rangle$} are conjugate and follow
the relation, $ |B_k\rangle = N^{-1/2} \sum_{l=0}^{N-1}\exp(2\pi ikl/N)|A_l\rangle \label{conj} $ The measurement
of the system by the probe under such a hamiltonian can be written as $ |s_j\rangle|A_k\rangle \rightarrow
|s_j\rangle |A_{k+j}\rangle \label{premsr}$. Of course the above representation for system probe interaction as a
closed system is not unique and entirely depends on the choice of basis for the system and probe. For an initial
probe state $|A_0\rangle$ and the combined (pure)state of system and probe $ |s_j\rangle|A_0\rangle = |s_j\rangle
N^{-1/2}\sum_{k=0}^{N-1} |B_k\rangle $. Any measurement is a process of the measurement device undergoing
decoherence. The probe which is a measurement device designed to monitor the system undergoes decoherence under
the probe environment interaction hamiltonian, $H_{PE}$ in a similar fashion. Since the probe is designed to
continuously and freely interact with the environment, it undergoes decoherence and collapses to a mixed state of
the pointer basis {$|A_i\rangle$} several orders of magnitude faster than the system, which gives us ample time to
observe the system before it succumbs to decoherence. Several research groups have carried out experiments to
monitor the decoherence of quantum systems by various measurement schemes eg. Rydberg atom system in a cavity by
Brune et.al,\cite{brune1}\cite{brune2} and spin squeezing and noise suppression in collective spin
systems\cite{geremia} using QND measurements and homodyne detection schemes. Our theory here falls well into the
framework of such measurements and the ensuing analysis encompasses such schemes as well. The time evolution under
the interaction hamiltonian, $H_{SP}$,
\begin{align*}
&e^{(-i\int H_{SP}dt/\hbar)}|s_j\rangle|A_0\rangle =|s_j\rangle N^{-1/2} \sum_{k=0}^{N-1}e^{(-ijk\int g(t)dt/\hbar)}|B_k\rangle\\
&= |s_j\rangle N^{-1}\sum_{l=0}^{N-1} \sum_{k=0}^{N-1}
e^{(2\pi ikl/N)} e^{(-ijk\int g(\tau)d\tau/\hbar)}|A_l\rangle\\
&= |s_j\rangle N^{-1} \sum_{k,l=0}^{N-1}\alpha_{jl}(t)|A_l\rangle
\end{align*}
where $ \alpha_{jl}(t) = \sum_k e^{(2\pi ik/N)(l-j\int_0^t g(\tau)d\tau/c\hbar)};c=\frac{2\pi}{N}$. The
pre-measurement\cite{zurek} is complete after a time when the integrated coupling strength $G=\int g(t)dt=\hbar
c$. The time scale of interaction is short compared to free evolution of the system or the environment and hence
can be assumed to be frozen for the period of interaction. Since the interaction of the probe is non-demolition
and only reveals information about the degree of decoherence we do not have to consider the dynamical evolution
due $H_{PE}$ in our system equation. Since the system and probe are entangled due to the interaction, any
measurement of the probe observable would also partially collapse the system into a statistical mixture of states.
Let the initial state of the system be $\sum c_j|s_j\rangle$ and the probe $|A_0\rangle$. The time evolution
yields,
\[
\sum_j c_j |s_j\rangle |A_0\rangle \rightarrow \sum_j
c_j|s_j\rangle\sum_{l=0}^{N-1} N^{-1} \alpha_{jl}(t) |A_l\rangle
\]
The theory of continuous measurement assumes that a successful measurement could be obtained even in the limit
as the interaction time goes to zero (i.e) $t\rightarrow 0$. And hence the superposition parameters
$\alpha_{jl}(t)$ are evaluated in the limiting case. The expected value of the probe observable (denoted
$\langle a(t) \rangle$) $\hat{A}$ is now,
\[
\langle a(t) \rangle=N^{-1}\sum_l a_l \sum_j|\alpha_{jl}(t)c_j|^2
\]
Following the same calculations it can be shown that the expected value would have been the same even if the
system were in a mixed state \{$p_j,|s_j\rangle$\} such that $|c_j|^2=p_j$. In order to surmount the problem it is
helpful to measure the system in the conjugate basis of the pointer basis. For a discrete N-level system the
conjugate basis states are given by, $ |s'_l\rangle = N^{-1/2} \sum_{k=0}^{N-1}e^{(2\pi ikl/N)}|s_k\rangle
\label{ex1} $ which is precisely the Quantum Fourier Transform (QFT) of the states \{$|s_k\rangle$\}. The mixed
state \{$p_j, |s_j\rangle$\} after the QFT would be transformed to $\{p_j,|s_j'\rangle\}$ where $|s_j'\rangle$ is
the QFT of the state $|s_j\rangle$. After the interaction with the probe for a time $t$, the system is now in the
state,
\[
N^{-3/2} \sum_{k=0}^{N-1}\exp\left(\frac{2\pi
i}{N}kj\right)|s_k\rangle\sum_{l=0}^{N-1}\alpha_{kl}|A_l\rangle
\]
with the same probability $p_j$. Hence the expected value of the continuous measurement for the mixed state in
the conjugate basis is,
\begin{align*}
\langle a(t) \rangle =&N^{-3/2}\sum_j p_j \sum_l
a_l\sum_k\left|\alpha_{kl}(t)\exp\left(\frac{2\pi i
kj}{N}\right)\right|^2 \\
=&N^{-3/2}\sum_l \sum_k a_l |\alpha_{kl}|^2
\end{align*}
The schematic for implementing such a measurement is shown in Fig. 1, where QFT is applied to the system before
and after each measurement, following the dotted pulse and the measurement is carried out during the complementary
solid pulse. Since the transform is unitary and self-adjoint the QFT applied just after the measurement is to
restore it back to its original state. A more general expression for measurement results for intermediary degrees
of {\it mixedness} can now be derived.
Consider the case of partial decoherence, in which case the joint state of system and environment can be written
as, $\xi(t)= \sum_{j=0}^{N-1} c_j|s_j\rangle|e_j(t)\rangle$, where the states of the environment correlated to the
system aren't quite orthogonal yet , i.e, $\langle e_i(t)|e_j(t)\rangle \neq \delta_{ij}$. It is convenient to
express the states $|e_j(t)\rangle$ in terms of the known classical states of the environment $|E_j\rangle$, which
needless to say are orthogonal to each other. So, $|e_j(t)\rangle = \sum_{j'=0}^{N-1} \lambda_{jj'}(t)
|E_{j'}\rangle$ with $\sum_{j'=0}^{N-1} |\lambda_{jj'}(t)|^2=1$. If the system exists in pure state, then
$|e_i(t)\rangle = |e_j(t)\rangle$, i.e, $ \lambda_{jj'}(t) = \lambda_{ij'}(t); \forall i,j $ and for a completely
decohered system in mixed state in the pointer basis $|s_j\rangle$, we have $\lambda_{jj'}(t)=\delta_{jj'}$, or in
general we require that $\langle e_i|e_j\rangle = \delta_{ij}$ which translates to $\sum_{j'}
\lambda^*_{ij'}(t).\lambda_{jj'}(t) = \delta_{ij}$. For ease of notation we will suppress the time index
henceforth and all the summations are assumed to be from $0$ to $N-1$ unless stated otherwise. If the state of the
joint system+environment is $\sum_j c_j|s_j\rangle|e_j\rangle$, then after the QFT and interaction with the
measurement device the state becomes,
\begin{align*}
N^{-3/2}\sum_j \sum_k c_j e^{(2\pi i jk/N)}|s_k\rangle.
\left(\sum_l \alpha_{kl} |A_l\rangle
\right).|e_j\rangle \\
=N^{-3/2}\sum_{j,k,l}c_j e^{(2\pi i jk/N)}\sum_{j'}\lambda_{jj'}
\alpha_{kl}|s_k\rangle|A_l\rangle|E_{j'}\rangle
\end{align*}
The expected value of a measurement performed on the apparatus
observable $\hat{A} = \sum_l a_l|A_l\rangle\langle A_l|$ is,
\begin{align*}
\langle a(t)\rangle = &N^{-3/2} \sum_l a_l \sum_{k,j'}\left|
\sum_j c_j e^{(2\pi i jk/N)}
\lambda_{jj'}\right|^2.|\alpha_{kl}|^2\\
= &N^{-3/2}\sum_{l,k,j'} a_l\left[ \sum_j|c_j|^2.|\lambda_{jj'}|^2+\right.\\
&\left.2\sum_{j,m} re[e^{(2\pi i k(j-m)/N)}c^*_jc_m. \lambda^*_{jj'}\lambda_{mj'}]\right].|\alpha_{kl}|^2
\end{align*}
$\forall j,m \mbox{, s.t }0\leq j,m \leq N-1$ and $m \neq j$. Two
simple scenarios can now be considered.,\\
 {\it Case I Pure State.} For a pure state the expected value
 just becomes(using properties of $\lambda_{jj'}$),
\begin{align*}
&N^{-3/2}\sum_{k,l} a_l (\sum_j |c_j|^2+ 2\sum_{j,m}
re\left[e^{(2\pi i k(j-m)/N)}c^*_jc_m\right])|\alpha_{kl}|^2\\
&=N^{-3/2}\sum_{l,k} a_l (1+ 2\sum_{j,m} r(j,m)).|\alpha_{kl}|^2;
\end{align*}
$\forall j \neq m$ where $r(j,m)=re\left[e^{(2\pi i
k(j-m)/N)}c^*_jc_m\right]$.\\
{\it Case II Mixed State}. The corresponding expected value for a
mixed state is,
\begin{align*}
&N^{-3/2}\sum_{l,k} a_l (\sum_j |c_j|^2+ 2\sum_{j,m}
r(j,m)\delta_{jm}).|\alpha_{kl}|^2;
j\neq m\\
&=N^{-3/2}\sum_{l,k} a_l |\alpha_{kl}|^2
\end{align*}
\begin{figure}
\epsfysize=1.5in \epsfxsize=2in \epsffile{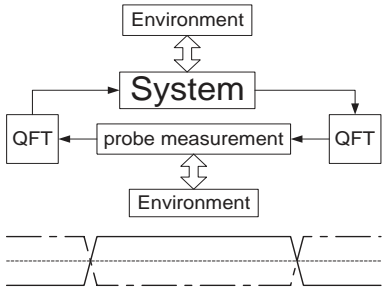} \caption{Schematic of a continuous measurement by probe
and pulse diagram showing activation of QFT and probe measurement respectively. The solid pulse represents system
probe interaction signal and dotted pulse which is complementary represents QFT.}
\end{figure}
    We may assume that the initial wave function of the probe before the interaction was $\phi(a)$ in the
$|A_k\rangle$ basis. The wave function of the probe after interaction with the quantum system becomes
$\phi(a+G\hat{s})$ where $G$ and $\hat{s}$ are as defined previously. This is due to the shift generated by the
interaction hamiltonian in the conjugate basis. The probability distribution of the probe measurement after the
system probe interaction is $f(a) = \sum_j |\phi(a+Gs_j)|^2|\langle s_j|\psi(t)\rangle|^2,$ where
$|\psi(t)\rangle$ is the combined state of system-probe\cite{opqusys}. Hence from the measurement result of the
probe observable $\hat{A}$, information about the system observable $\hat{s}$ is given by the probability
distribution, $ f(s) = \sum_j W(s-s_j)|\langle s_j|\psi(t)\rangle|^2, $ where $W(s-s_j) = |G|.|\phi(\langle a
\rangle-G(s-s_j))|^2$. Results from standard quantum tomography can now be applied to the probability distribution
of $\hat{s}$ to obtain the estimate of the state in least square sense\cite{buzek}. Hence it is possible to get an
estimate of the type of state as well as a reasonably good estimate of the state itself from the expected value of
the probe measurement which in turn is used to design the state feedback discussed above\cite{ganesan2}. Such a
problem at hand would invoke techniques from non-linear output regulation\cite{huangjie} wherein the control
inputs could be designed to steer the output (\ref{output}) away from or towards a desired output value.
    This research was supported in part by the U. S. Army Research Office under Grant W911NF-04-1-0386. T. J. Tarn
would also like to acknowledge partial support from the China Natural Science Foundation under Grant Number
60433050 and 60274025.

\end{document}